# Solitary, explosive, rational and elliptic doubly periodic solutions for nonlinear electron-acoustic waves in the earth's magnetotail region

S. A. El-Wakil, E. M. Abulwafa, E. K. El-Shewy, H. M. Abd-El-Hamid

*Theoretical Physics Group, Physics Department, Faculty of Science, Mansoura University, Mansoura, Egypt*

**Abstract:** A theoretical investigation has been made of electron acoustic wave propagating in unmagnetized collisionless plasma consisting of a cold electron fluid and isothermal ions with two different temperatures obeying Boltzmann type distributions. Based on the pseudo-potential approach, large amplitude potential structures and the existence of Solitary waves are discussed. The reductive perturbation method has been employed to derive the Korteweg-de Vries (KdV) equation for small but finite amplitude electrostatic waves. An algebraic method with computerized symbolic computation, which greatly exceeds the applicability of the existing tanh, extended tanh methods in obtaining a series of exact solutions of the KdV equation, is used here. Numerical studies have been made using plasma parameters close to those values corresponding to Earth's plasma sheet boundary layer region reveals different solutions i.e., bell-shaped solitary pulses and singularity solutions at a finite point which called "blowup" solutions, Jacobi elliptic doubly periodic wave, a Weierstrass elliptic doubly periodic type solutions, in addition to the propagation of an explosive pulses. The result of the present investigation may be applicable to some plasma environments, such as earth's magnetotail region and terrestrial magnetosphere.

## 1. Introduction

The nonlinear evolution of electrostatic waves in plasmas is an important topic in plasma physics. Electron acoustic waves (EAWs) have been observed in the laboratory when the plasma consisted of two species of electrons with different temperatures, referred to as hot and cold electrons [1-2], or in an electron ion plasma with ions hotter than electrons [3]. Also its propagation plays an important role not only in laboratory but also in space plasma.

For example, bursts of broadband electrostatic noise (BEN) emissions have been observed in auroral and other regions of the magnetosphere, e.g. polar cusp, plasma sheet boundary layer (PSBL), see for instances [4-6]. There have been numerous observations of small temporal and spatial scale, large amplitude electric fields in the Earth's magnetotail region. These structures have been commonly called solitary waves or weak doubles layers and appear to be prevalent throughout many parts of the Earth's magnetosphere [7-8].

Investigations of small-amplitude electron acoustic waves (EAWs) in collisionless plasma usually describe the evolution of the wave by nonlinear equations such as the Korteweg-de Vries (KdV), KdV-type, Zakharov-Kuznetsov (ZK), ZK-type and nonlinear Schrödinger equations [9-13].

In the past several decades, new exact solutions may help to find new phenomena. A variety of powerful methods, such as inverse scattering method [13], bilinear transformation [15], the tanh-sech method [16], extended tanh method [17, 18], homogeneous balance method [19], exp-function method [20] and variational iteration method [21]. Fan [22] developed a new algebraic method with computerized symbolic computation, which greatly exceeds the applicability of the existing tanh, extended tanh methods and Jacobi function expansion method in obtaining a series of exact solutions of nonlinear equations.

The major topic of this work is to study the existence of the electrostatic arbitrary and small amplitude solitary and other type waves in unmagnetized collisionless plasma consists of a cold electron fluid and isothermal ions with two different temperatures obeying Boltzmann type distributions.

This paper is organized as follows: in section 2, we present the basic set of fluid equations governing our plasma model. In section 3, the pseudo-potential was discussed. In section 4, the nonlinear Small-Amplitude EAWs are investigated through the derivation of a Korteweg-de Vries (KdV) equation. In Section 5, an algorithm describing the computerized symbolic computation method is presented. In Section 6, the proposed method is applied to the KdV equation. Section 7 contains the results and discussions. Some conclusions are given in section 8.

## 2. Basic Equations

We consider a homogeneous system of unmagnetized collisionless plasma consists of a cold electron fluid and isothermal ions with two different temperatures (low-temperature $T_l$ and high-temperature $T_h$) obeying Boltzmann type distributions. Such system is governed, in one dimension, by the following normalized equations [23]:

$$\frac{\partial}{\partial t} n_e(x,t) + \frac{\partial}{\partial x}[n_e(x,t) u_e(x,t)] = 0, \qquad (1a)$$

$$\frac{\partial}{\partial t} u_e(x,t) + u_e(x,t) \frac{\partial}{\partial x} u_e(x,t) - \frac{\partial}{\partial x} \phi(x,t) = 0, \qquad (1b)$$

$$\frac{\partial^2}{\partial x^2} \phi(x,t) - n_e(x,t) + n_{il}(x,t) + n_{ih}(x,t) = 0. \qquad (1c)$$

In the above equations, $n_e(x,t)$ is the cold electron density normalized by initial equilibrium electrons density $n_{e0}$, $u_e(x,t)$ is the electron fluid velocity normalized by the effective electron acoustic speed $C_e = \sqrt{\dfrac{T_{eff}}{m_e}}$ where $T_{eff} = \dfrac{T_l T_h}{n_{ih0} T_l + n_{il0} T_h}$, $n_{il0}$ and $n_{ih0}$ are the initial normalized equilibrium densities of the low- and high-temperature ions, respectively with $n_{il0} + n_{ih0} = 1$, $\phi(x,t)$ is the electrostatic potential normalized by the effective electrostatic potential $\dfrac{T_{eff}}{e}$, where $m_e$ and $e$ are the mass and charge of the electron, $x$ is the space coordinate normalized by the effective Debye length $\lambda_{Deff} = \sqrt{\dfrac{T_{eff}}{4\pi n_{e0} e^2}}$ and $t$ is the time variable normalized by the inverse of the cold electron plasma frequency $\omega_{ce}^{-1}$ ($\omega_{ce} = \sqrt{\dfrac{4\pi n_{e0} e^2}{m_e}}$).

Equations (1a) and (1b) represent the inertia of the electron fluid and (1c) is the Poisson's equation needed to make the self-consistent. The two types of ions are described with distributions given by:

$$n_{il}(x,t) = n_{il0} \exp[-\dfrac{\phi(x,t)}{(n_{il0} + n_{ih0}\beta)}], \qquad (2a)$$

and

$$n_{ih}(x,t) = n_{ih0} \exp[-\dfrac{\beta \phi(x,t)}{(n_{il0} + n_{ih0}\beta)}], \qquad (2b)$$

where $\beta = T_l/T_h$ is the ions temperature ratio.

## 3. Nonlinear arbitrary amplitude

To investigate the nonlinear properties of the electrostatic waves described in section 2, we must consider the nonlinear terms in (1) and (2). Therefore, in order to study the fully nonlinear (arbitrary amplitude solitary) waves, we employ here the pseudo-potential approach [24] by assuming that all dependent variables depend on a single variable $\zeta = x - M t$, where $M$ is the Mach number (solitary wave speed normalized by the effective electron acoustic speed $C_e$). Using this transformation along with the steady state condition ($\dfrac{\partial}{\partial t} = 0$) and appropriate boundary conditions for localized distributions (as $|\zeta| \to \infty$, $n_e = 1$, $u_e = 0$ and $\phi = 0$), (1) and (2) can be reduced to

$$\dfrac{1}{2}[\dfrac{d\phi}{d\zeta}]^2 + V(\phi) = 0, \qquad (3a)$$

where the potential $V(\phi)$ is given by

$$V(\phi) = M(\sqrt{M^2} - \sqrt{M^2 + 2\phi}) + \dfrac{n_{ih0}}{\beta}(n_{il0} + n_{ih0}\beta)[1 - \exp(-\dfrac{\beta\phi}{n_{il0} + n_{ih0}\beta})]$$

$$+ n_{il0}(n_{il0} + n_{ih0}\beta)[1 - \exp(-\frac{\phi}{n_{il0} + n_{ih0}\beta})]. \quad (3b)$$

The nonlinear equation (3a) can be regarded as "energy integral" of an oscillating particle of unit mass with velocity $\frac{d\phi}{d\zeta}$ and position $\phi$. This equation is valid for arbitrary amplitude electron-acoustic waves in the steady state. A necessary conditions for the existence of the solitary waves is that $V(\phi = 0) = 0$, $\left.\frac{\partial V}{\partial \phi}\right|_{\phi=0} = 0$, and $\frac{\partial^2 V}{\partial \phi^2} < 0$ and $V(\phi) < 0$ for $\phi$ lying between 0 and $\phi_m$.

## 4. Nonlinear small-amplitude

According to the general method of reductive perturbation theory, slow stretched coordinates are introduced as [25]:

$$\tau = \varepsilon^{3/2} t, \text{ and } \xi = \varepsilon^{1/2}(x - vt), \quad (4)$$

where $\varepsilon$ is a small dimensionless expansion parameter and $v$ is the wave speed normalized by $C_e$. All physical quantities appearing in (1) are expanded as power series in $\varepsilon$ about their equilibrium values as:

$$n_c = 1 + \varepsilon n_1 + \varepsilon^2 n_2 + \varepsilon^3 n_3 + ..., \quad (5a)$$

$$u_c = \varepsilon u_1 + \varepsilon^2 u_2 + \varepsilon^3 u_3 + ..., \quad (5b)$$

$$\phi = \varepsilon \phi_1 + \varepsilon^2 \phi_2 + \varepsilon^3 \phi_3 + .... \quad (5c)$$

We impose the boundary conditions as $|\xi| \to \infty$, $n_e = 1$, $u_e = 0$ and $\phi = 0$.

Substituting (4) and (5) into (1) and equating coefficients of like powers of $\varepsilon$ lead, from the two lowest-order equations in $\varepsilon$, to the following KdV equation for the first-order perturbed potential:

$$\frac{\partial}{\partial \tau}\phi_1(\xi,\tau) + A\phi_1(\xi,\tau)\frac{\partial}{\partial \xi}\phi_1(\xi,\tau) + B\frac{\partial^3}{\partial \xi^3}\phi_1(\xi,\tau) = 0, \quad (6a)$$

where

$$A = \frac{v^3}{2}[\frac{(n_{il0} + n_{ih0}\beta^2)}{(n_{il0} + n_{ih0}\beta)^2} - \frac{3}{v^4}], \quad B = \frac{v^3}{2} \quad \text{and} \quad v = \pm 1. \quad (6b)$$

This KdV equation can be solved using a computerized symbolic computational technique [22].

## 5. Computerized symbolic computation method

Fan developed a computerized symbolic computation method to solve nonlinear partial differential equation. This technique can be represented as follows [22]:

**Step 1:** For a given partial differential equation in $\phi_1(x,t)$ of the form

$$H(\phi_1, \frac{\partial \phi_1}{\partial \tau}, \frac{\partial \phi_1}{\partial \xi}, \frac{\partial^2 \phi_1}{\partial \xi^2}, ...) = 0. \tag{7}$$

In a traveling frame of reference $\phi_1(\xi, \tau) = \Phi(\eta)$, $\eta = \xi - \Lambda \tau$, the partial differential equation may be transformed into an ordinary differential equation of the form

$$H(\Phi, \frac{d\Phi}{d\eta}, \frac{d^2\Phi}{d\eta^2}, ...) = 0. \tag{8}$$

**Step 2:** Expand the solution of (8) in the form

$$\Phi(\eta) = \sum_{i=0}^{n} a_i \, \varphi(\eta)^i, \tag{9}$$

where $a_i$, ($i = 0, 1, ..., n$) are coefficients to be determined and the new variable $\varphi(\eta)$ is a solution of the following ordinary differential equation

$$\frac{d\varphi}{d\eta} = \pm \sqrt{\sum_{j=0}^{k} c_j \varphi(\eta)^j}, \tag{10}$$

where $c_j$, ($j = 0, 1, ..., k$) are coefficients will be determined.

**Step 3:** Substituting (9) and (10) into (8) and balancing the highest derivative term with the nonlinear term lead to a relation between $n$ and $k$, from which the different possible values of $n$ and $k$ can be obtained. These values lead to the different series expansions of the exact solutions of the nonlinear differential equation (8).

**Step 4:** Substituting (9) and (10) into (8) and setting the coefficients of all powers of $\varphi(\eta)^i$ and $\varphi(\eta)^i \frac{d\varphi}{d\eta}$ equal to zero will give a system of algebraic equations. From this set of algebraic equations, the parameters $\Lambda$, $a_i$, ($i = 0, 1, ..., n$) and $c_j$, ($j = 0, 1, ..., k$) can be found explicitly.

**Step 5:** Substituting the parameters $c_j$ ($j = 0, 1, ..., k$) obtained in step 4 into (10), we can then obtain all the possible solutions of $\varphi(\eta)$. We remark here that the solutions of (8) depend on the explicit solvability of (10). The solution of the system of algebraic equations will give a series of fundamental solutions such as polynomial, exponential, soliton, rational and triangular periodic solutions.

## 6. Explicit exact solutions for the KdV equation

For KdV equation (6a), the traveling wave transformation $\phi_1(\xi, \tau) = \Phi(\eta)$, $\eta = \xi - \Lambda \tau$, gives an ordinary differential equation of the form

$$-\Lambda \frac{d}{d\eta}\Phi(\eta) + A\Phi(\eta)\frac{d}{d\eta}\Phi(\eta) + B\frac{d^3}{d\eta^3}\Phi(\eta) = 0, \tag{11}$$

where the coefficients $A$ and $B$ are given by (6b) and $\Lambda$ is an arbitrary parameter similar to the Mach number ($M$).

Substituting the solution given by (9) and (10) and balancing the highest nonlinear term with the highest derivative term in resultant equation give the relation between $n$ and $k$ as $k = n + 2$. If we take $n = 2$ then $k = 4$, the solution of (11) can be represented as

$$\Phi(\eta) = a_0 + a_1 \varphi(\eta) + a_2 \varphi(\eta)^2, \tag{12}$$

$$\frac{d\varphi}{d\eta} = \pm \sqrt{c_0 + c_1 \varphi(\eta) + c_2 \varphi(\eta)^2 + c_3 \varphi(\eta)3 + c_4 \varphi(\eta)4}. \tag{13}$$

Using the symbolic software package Maple, we obtain the following solutions for the KdV equation (11):

$$\Phi(\eta) = a_0 + 3\frac{B}{A} Y \operatorname{sech}(\frac{1}{2}\sqrt{Y}\eta)^2, \tag{14}$$

$$\Phi(\eta) = a_0 + 3\frac{B}{A} Y \sec(\frac{1}{2}\sqrt{-Y}\eta)^2, \tag{15}$$

$$\Phi(\eta) = a_0 + \frac{3}{2}\frac{B}{A} Y \tanh(\frac{1}{2}\sqrt{\frac{-Y}{2}}\eta)^2, \tag{16}$$

$$\Phi(\eta) = a_0 - \frac{3}{2}\frac{B}{A} Y \tan(\frac{1}{2}\sqrt{\frac{Y}{2}}\eta)^2, \tag{17}$$

$$\Phi(\eta) = a_0 - 3\frac{B}{A} Y \operatorname{csch}(\frac{1}{2}\sqrt{Y}\eta)^2, \tag{18}$$

$$\Phi(\eta) = a_0 + 3\frac{B}{A} Y \csc(\frac{1}{2}\sqrt{-Y}\eta)^2, \tag{19}$$

$$\Phi(\eta) = \frac{\Lambda}{A} - \frac{12 B}{A \eta^2}, \tag{20}$$

$$\Phi(\eta) = a_0 + 3\frac{B}{A} Y \frac{m^2}{2m^2 - 1} cn(\frac{1}{2}\sqrt{\frac{-Y}{(1 - 2m^2)}}\eta)^2, \tag{21}$$

$$\Phi(\eta) = a_0 - 3\frac{B}{A} Y \frac{1}{m^2 - 2} dn(\frac{1}{2}\sqrt{\frac{-Y}{(m^2 - 2)}}\eta)^2, \tag{22}$$

$$\Phi(\eta) = a_0 - 3\frac{B}{A} Y \frac{m^2}{m^2 + 1} sn(\frac{1}{2}\sqrt{\frac{-Y}{(m^2 + 1)}}\eta)^2, \tag{23}$$

$$\Phi(\eta) = a_0 - 3(B/A)(2c_3)^{2/3} \wp(C\eta; g_2, g_3), \tag{24a}$$

$$C = (c_3/4)^{1/3}, \quad g_2 = -c_1/C \quad \text{and} \quad g_3 = -c_0, \tag{24b}$$

where $Y = (\Lambda - a_0 A)/B$, $a_0$, $c_0$, $c_1$ and $c_3$ are arbitrary constants, $m$ is a modulus and $\wp(C\eta; g_2, g_3)$ is Weierstrass function with invariants $g_2$ and $g_3$.

## 7. Results and discussion

To make our result physically relevant, numerical studies have been made using plasma parameters close to those values corresponding to earth's plasma sheet boundary layer region [8, 26].

However, since one of our motivations was to study the effects of Mach number ($M$), the initial normalized low temperature ions density ($n_{il0}$) and the ions temperature ratio ($\beta$) on the existence of electrostatic solitary waves by analyzing the Sagdeev's pseudo-potential $V(\phi)$ for arbitrary amplitude electron-acoustic waves. For example, Figures 1, 2 and 3 concern the effects of $M$, $n_{il0}$ and $\beta$, respectively on the existence of the electrostatic solitary waves.

In Fig. (1), the behavior of $V(\phi)$ shows the critical Mach number ($M$) for which a potential well in the negative-axis (corresponding to a solitary wave with a negative potential) develops. Increasing the value of $M$ increases the negative depth and the width of the potential well.

The variation of $V(\phi)$ with $n_{il0}$ is represented in Fig. (2). This figure shows that initial normalized low temperature ions density has nearly the same effect on the Sagdeev's potential as the Mach number.

Figure (3) shows the effect of changing the parameter $\beta$ on the potential $V(\phi)$. Increasing the $\beta$-value decreases both the negative depth and the width of Sagdeev's potential $V(\phi)$.

For small-amplitude electron-acoustic waves, the Korteweg-de Vries equation has been derived using the reductive perturbation method. A symbolic computational traveling wave method is used to obtain a series of exact solutions of KdV equation [22].

The solutions given by (14) and (15) tend, as $a_0 = 0$, to the stationary solution of this problem given by Kakad et al [21]. In Figs. (4) and (5), a profile of the bell-shaped compressive and rarefactive solitary waves for the electrostatic potentials and the associated bipolar electric field structures are obtained for solutions (14) and (15).

Figure (4) shows the variation of the electrostatic solitary waves from compressive to rarefactive forms due to the variation of the initial normalized low-temperature ions density ($n_{il0}$). The change from compressive to rarefactive solitary waves occurs at about $n_{il0} \approx 0.25594$ for $\beta$=0.05, $\Lambda$=0.001, $v$=1 and $a_0$=0. The increasing of the parameter $n_{il0}$ increases both the amplitude and width of the compressive solitary wave while it decreases the values of the amplitude and the width of the rarefactive waves.

The effect of the ions temperature ratio ($\beta$) on the electrostatic solitary waves is represented in Fig. (5). The parameter ($\beta$) has similar effect as the parameter $n_{il0}$ on the variation of the solitary waves from compressive to rarefactive solitons and also on the variation of the waves amplitude and width. The value of $\beta$, which the electrostatic solitary waves change from compressive to rarefactive, is about 0.063296 for $n_{il0}$=0.23, $\Lambda$=0.001, $v$=1 and $a_0$=0.

Solutions (16) and (17) develop solitons with singularity at a finite point which called "blowup" of solutions. The profile of doubly rarefactive and compressive solitary pulses of blowup points are depicted in Fig. (6) for solutions (16) and (17).

On the other hand, solutions (18), (19) and the rational solution (20) lead to the propagation of explosive solitary pulse. The explosive rarefactive and compressive electrostatic solitary waves due to the solutions (18)-(20) are represented in Fig. (7).

Solutions (21)-(23) are three Jacobi elliptic functions for wave solutions. When $m \to 1$, the Jacobi elliptic functions (21)-(23) degenerate to the hyperbolic functions (14), (16) and (18) while as $m \to 0$ they degenerate to the triangular functions (15), (17) and (19).

Finally, solution (24) gives a Weierstrass elliptic doubly periodic type solution. The profile of the Weierstrass elliptic doubly periodic type solution and the associated bipolar electric field structures are shown in Fig. (8).

In summery, it has been found that the presence of two isothermal ions with different temperatures modifies the properties of the electron acoustic solitary waves significantly and new solutions have been obtained. To our knowledge, these solutions have not been reported. It may be important to explain some physical phenomena in some plasma environments, such as earth's plasma sheet boundary layer region.

## 8. Conclusion

We have devoted quite some efforts to discuss the proper description of new solutions in unmagnetized collisionless plasma consisting of a cold electron fluid and isothermal ions with two different temperatures obeying Boltzmann type distributions. The application of the pseudo-potential approach leads to Sagdeev's pseudo-potential form for arbitrary amplitude electron-acoustic waves. It emphasizes the amplitude of the electron-acoustic waves as well as parametric regime where the solitons can exist is sensitive to the Mach number ($M$), the low temperature ion density ($n_{il0}$) and the ions temperature ratio ($\beta$).

For small amplitude electron-acoustic waves, the study of the reductive perturbation theory gives the KdV equation. Generally speaking, our model admits the coexistence of both rarefactive and compressive solitons and other types of solutions. Moreover, new exact solutions to the KdV equation provide guidelines to classify the types of solutions according to the plasma parameters and can admits the following types of solutions: (a) hyperbolic and solitary wave solutions, (b) triangular periodic wave solutions, (c) rational solutions, (d) Jacobi elliptic doubly periodic wave solutions and (e) Weierstrass elliptic doubly periodic type solution. The application of our model might be particularly interesting in the new observations for the earth's plasma sheet boundary layer region.

# References

[1] H. Derfler, T. C. Simonen, Higher-Order Landau Modes, *Physics of Fluids* **12(2)**, 269-287 (1969).

[2] S. Ikezawa and Y. Nakamura, Observation of Electron Plasma Waves in Plasma of Two-Temperature Electrons, *J. Physical Society of Japan* **50(3)**, 962-967 (1981).

[3] B. D. Fried and R. W. Gould, Longitudinal Ion Oscillations in a Hot Plasma, *Physics of Fluids* **4(1)**, 139-147 (1961).

[4] F. S. Mozer, R. Ergun, M. Temerin, C. A. Cattell, J. Dombeck and J. Wygant, New Features of Time Domain Electric-Field Structures in the Auroral Acceleration Region, *Physical Review Letters* **79(7)**, 1281-1284 (1997).

[5] R. E. Ergun, C. W. Carlson, J. P. McFadden, F. S. Mozer, G. T. Delory, W. Peria, C. C. Chaston, M. Temerin, I. Roth, L. Muschietti, R. Elphic, R. Strangeway, R. Pfaff, C. A. Cattell, D. Klumpar, E. Shelley, W. Peterson, E. Moebius and L. Kistler, FAST satellite observations of large-amplitude solitary structures, *Geophysical Research Letters* **25(12)**, 2041-2044 (1998).

[6] R. Pottelette, R. E. Ergun, R. A. Treumann, M. Berthomier, C. W. Carlson, J. P. McFadden and I. Roth, Modulated electron-acoustic waves in auroral density cavities: FAST observations, *Geophysical Research Letters* **26(16)**, 2629-2632 (1999).

[7] C. A. Cattell, J. Dombeck, J. R. Wygant, M. K. Hudson, F. S. Mozer, M. A. Temerin, W. K. Peterson, C. A. Kletzing, C. T. Russell and R. F. Pfaff, Comparisons of Polar satellite observations of solitary wave velocities in the plasma sheet boundary and the high altitude cusp to those in the auroral zone, *Geophysical Research Letters* **26(3)**, 425-428 (1999).

[8] H. Matsumoto, H. Kojima, T. Miyatake, Y. Omura, M. Okada, I. Nagano and M. Tsutui, Electrostatic solitary waves (ESW) in the magnetotail: BEN wave forms observed by GEOTAIL, *Geophysical Research Letters* **21(27)**, 2915-2918 (1994).

[9] S. G. Tagare, S. V. Singh, R. V. Reddy and G. S. Lakhina, Electron acoustic solitons in the Earth's magnetotail, *Nonlinear Processes in Geophysics* **11(2)**, 215-218 (2004).

[10] E. K. El-Shewy, Linear and nonlinear properties of electron-acoustic solitary waves with non-thermal electrons, *Chaos, Solitons and Fractals* **31(4)**, 1020-1024 (2007).

[11] S. A. Elwakil, M. A. Zahran and E. K. El-Shewy, Nonlinear electron-acoustic solitary waves in a relativistic electron-beam plasma system with non-thermal electrons, *Physica Scripta* **75(5)**, 803-808 (2007).

[12] I. Kourakis and P. K. Shukla, Electron-acoustic plasma waves: Oblique modulation and envelope solitons, *Physical Review E* **69(3)**, 036411 (2004).

[13] P. K. Shukla, L. Stenflo and M. Hellberg, Dynamics of coupled light waves and electron-acoustic waves, *Physical Review E* **66(2)**, 027403 (2002).

[14] M. J. Ablowitz and P. A. Clarkson, "*Solitons, Nonlinear evolution equations and Inverse scattering*", (Cambridge University Press, Cambridge, 1991).

[15] R. Hirota and J. Satsuma, Soliton solutions of a coupled Korteweg-de Vries equation, *Physics Letters A* **85(8-9)**, 407-408 (1981).

[16] W. Malfliet, Solitary wave solutions of nonlinear wave equations, *American Journal of Physics* **60(7)**, 650-654 (1992).

[17] E. Fan, Extended tanh-function method and its applications to nonlinear equations, *Physics Letters A* **277(4-5)**, 212-218 (2000).

[18] S. A. Elwakil, S. K. El-Labany, M. A. Zahran and R. Sabry, Modified extended tanh-function method for solving nonlinear partial differential equations, *Physics Letters A* **299(2-3)**, 179-188 (2002).

[19] M. Wang, Y. Zhou and Z. Li, Application of homogeneous balance method to exact solutions of nonlinear equations in mathematical physics, *Physics Letters A* **216(1-5)**, 67-75 (1996).

[20] J. H. He an M. A. Abdou, New periodic solutions for nonlinear evolution equations using Exp-function method, *Chaos, Solitons and Fractals* **34(5)**, 1421-1429 (2007).

[21] J. H. He, Variational iteration method for autonomous ordinary differential systems, *Applied Mathematics and Computation* **114(2-3)**, 115-123 (2000).

[22] E. Fan, Uniformly constructing a series of explicit exact solutions to nonlinear equations in mathematical physics, *Chaos, Solitons and Fractals* **16(6)**, 819-839 (2003).

[23] A. P. Kakad, S. V. Singh, R. V. Reddy, G. S. Lakhina and G. S. Tagare, Electron acoustic solitary waves in the Earth's magnetotail region, *Advances in Space Research* **43(12)**, 1945-1949 (2009).

[24] R. Z. Sagdeev, Cooperative phenomena and shock waves in collisionless plasmas, in "***Review of Plasma Physics***", edited by M. A. Leontovich, vol 4, p 23, (Consultants Bureau: 1966).

[25] H. Washimi and T. Taniuti, Propagation of Ion-Acoustic Solitary Waves of Small Amplitude, *Physical Review Letters* **17(19)**, 996-998 (1966).

[26] S. V. Singh and G. S. Lakhina, Generation of electron-acoustic waves in the magnetosphere, *Planetary and Space Science* **49(1)**, 107-114 (2001).

**Figure Captions**

Fig (1): The behavior of Sagdeev's potential $V(\phi)$ vs $\phi$ with $n_{il0}=0.25$ and $\beta=0.07$ for different values of $M$.

Fig (2): The behavior of Sagdeev's potential $V(\phi)$ vs $\phi$ with $M=1.10$ and $\beta=0.07$ for different values of $n_{il0}$.

Fig (3): The behavior of Sagdeev's potential $V(\phi)$ vs $\phi$ with $n_{il0}=0.25$ and $M=1.10$ for different values of $\beta$.

Fig (4): A bell-shaped solitary pulse represented by the solution (14) shows the variation of the amplitude and width with $n_{il0}$ for $\beta=0.05$, $\Lambda=0.001$, $v=1$ and $a_0=0$: (a) electrostatic potential and (b) the associated bipolar electric field structures.

Fig (5): A bell-shaped solitary pulse represented by the solution (14) shows the variation of the amplitude and width with $\beta$ for $n_{il0}=0.23$, $\Lambda=0.001$, $v=1$ and $a_0=0$: (a) electrostatic potential and (b) the associated bipolar electric field structures.

Fig (6): A periodic pulse represented by solution (16) shows the variation of the amplitude of the electrostatic potential wave with $\beta$ for $n_{il0}=0.23$, $\Lambda=0.001$, $v=1$ and $a_0=0$: (a) rarefactive and (b) compressive

Fig (7): An explosive pulse represented by solution (18) shows the variation of the amplitude of the electrostatic potential wave with $\beta$ for $n_{il0}=0.23$, $\Lambda=0.001$, $v=1$ and $a_0=0$: (a) rarefactive and (b) compressive.

Fig (8): The profile of Weierstrass elliptic doubly periodic type solution (26) for $\beta=0.05$, $\Lambda=0.001$, $v=1$, $c_0=-0.1$, $c_1=1$, $c_3=0.01$ and $a_0=0.5:a_0=0$: (a) electrostatic potential and (b) the associated bipolar electric field structures.

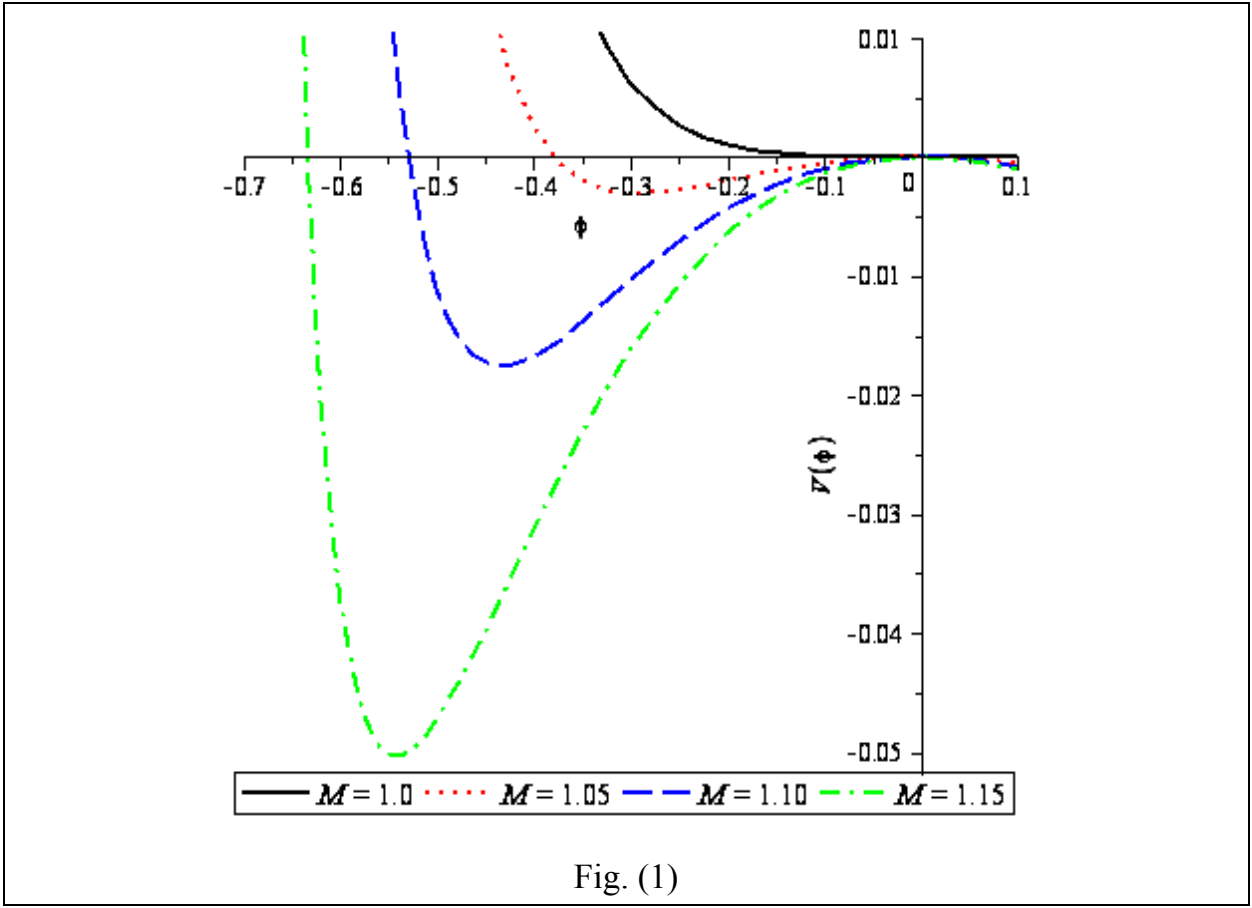

Fig. (1)

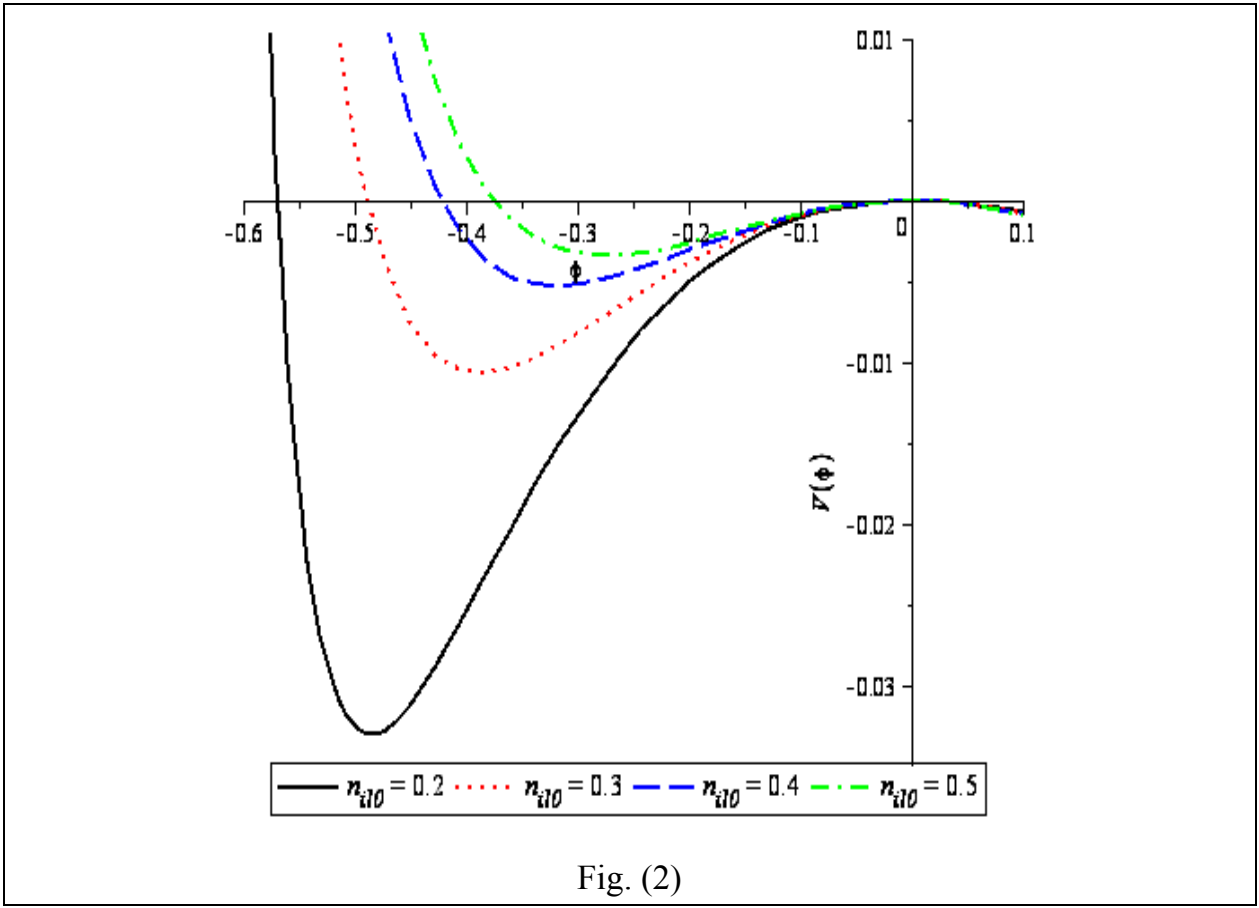

Fig. (2)

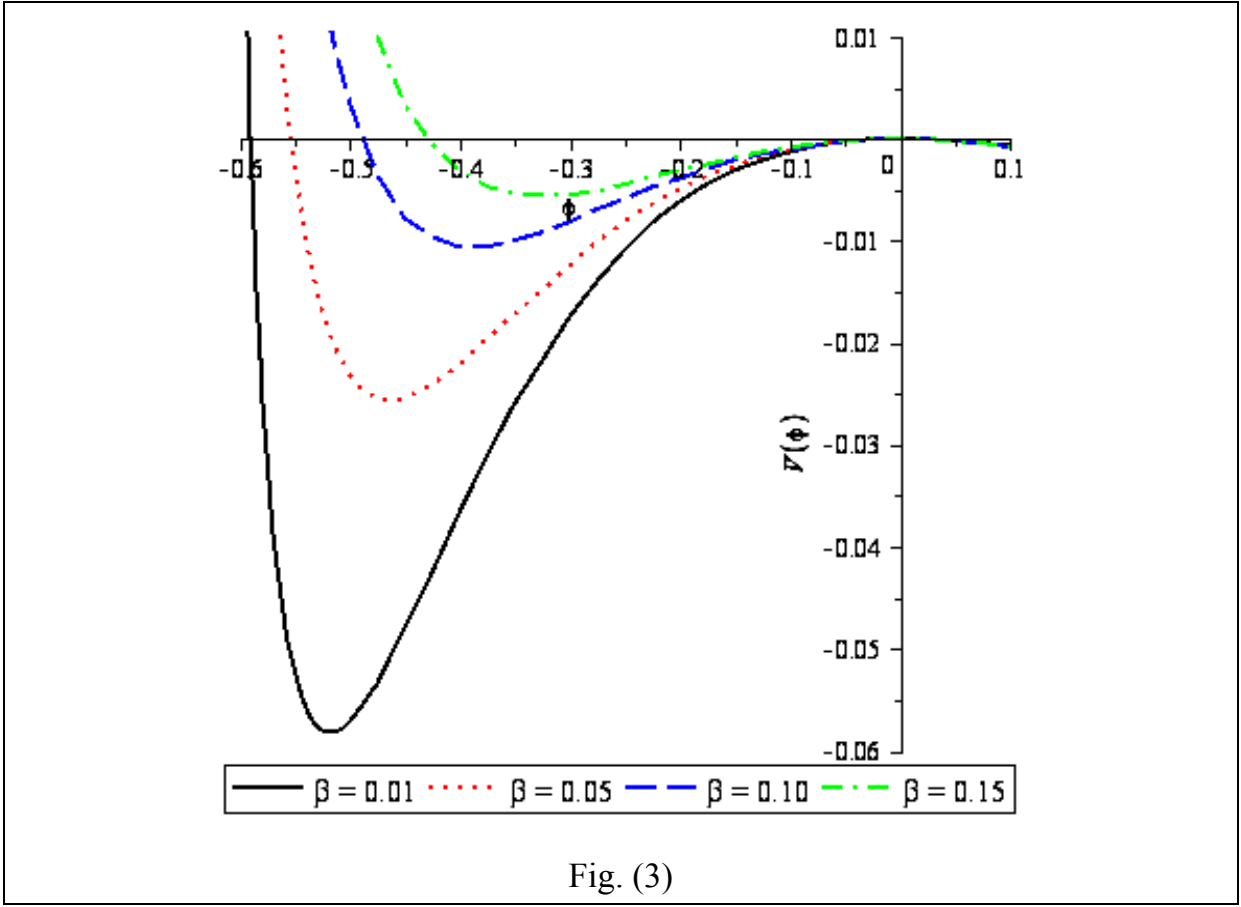

Fig. (3)

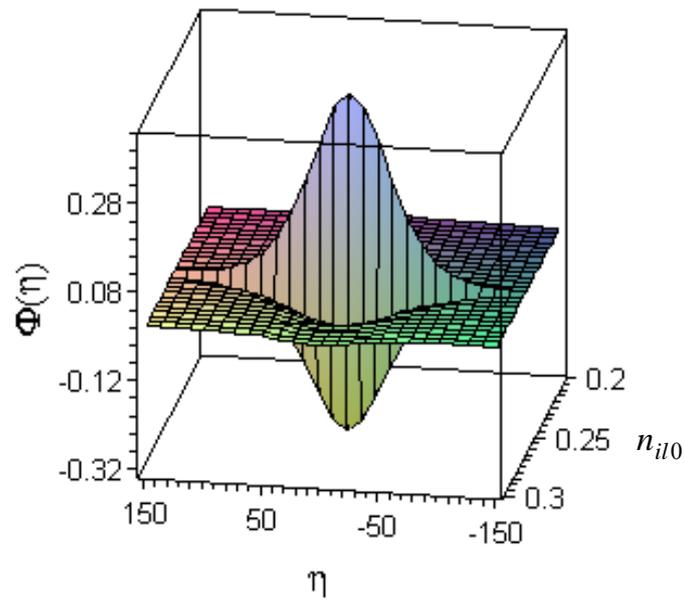

(a)

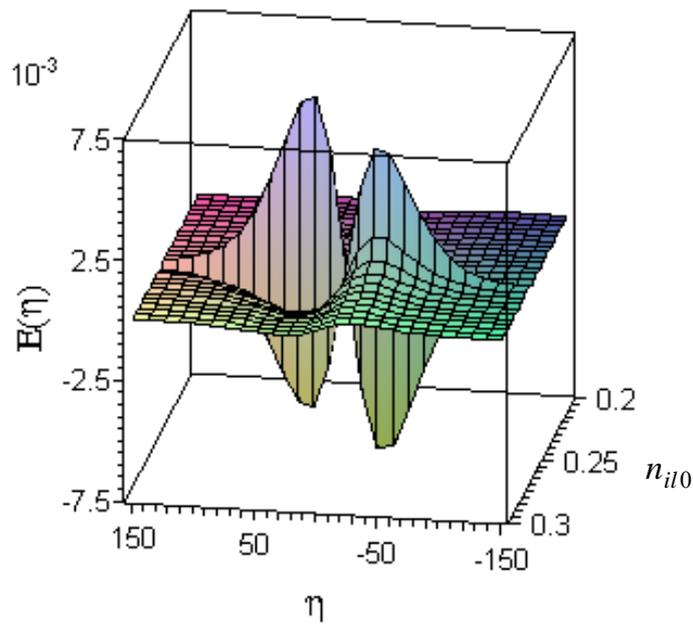

(b)

Fig. (4)

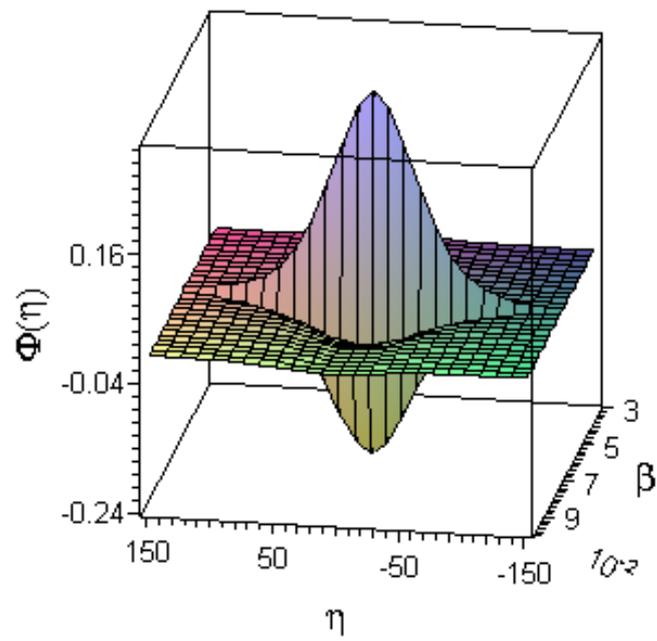

(a)

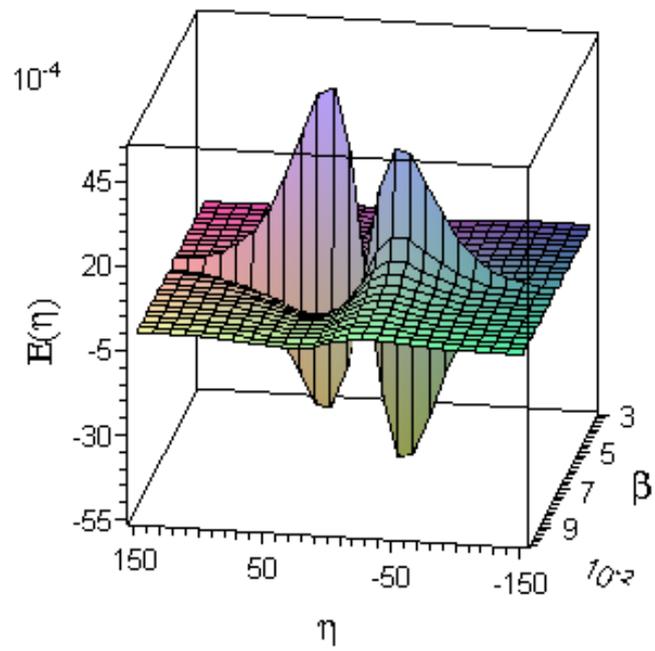

(b)

Fig. (5)

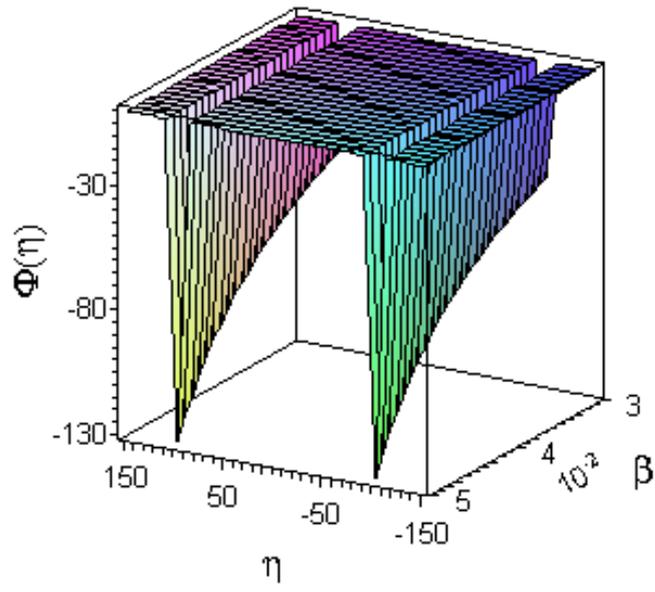

(a)

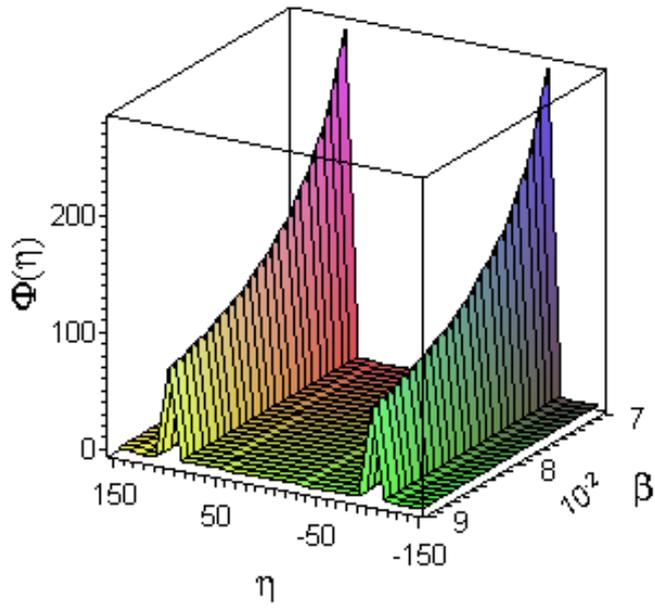

(b)

Fig. (6)

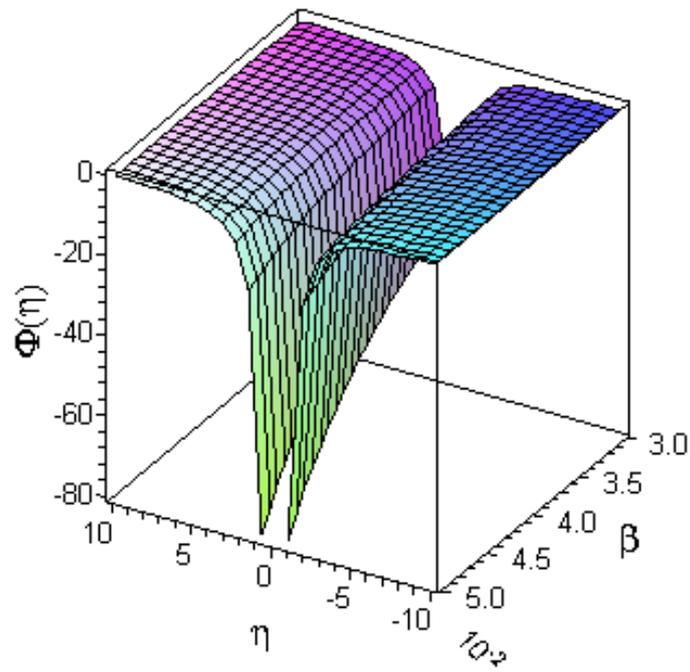

(a)

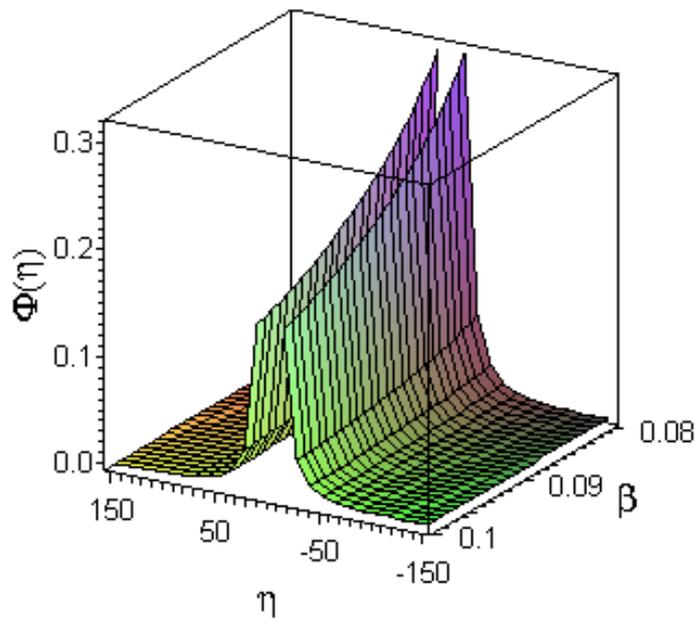

(b)

Fig. (7)

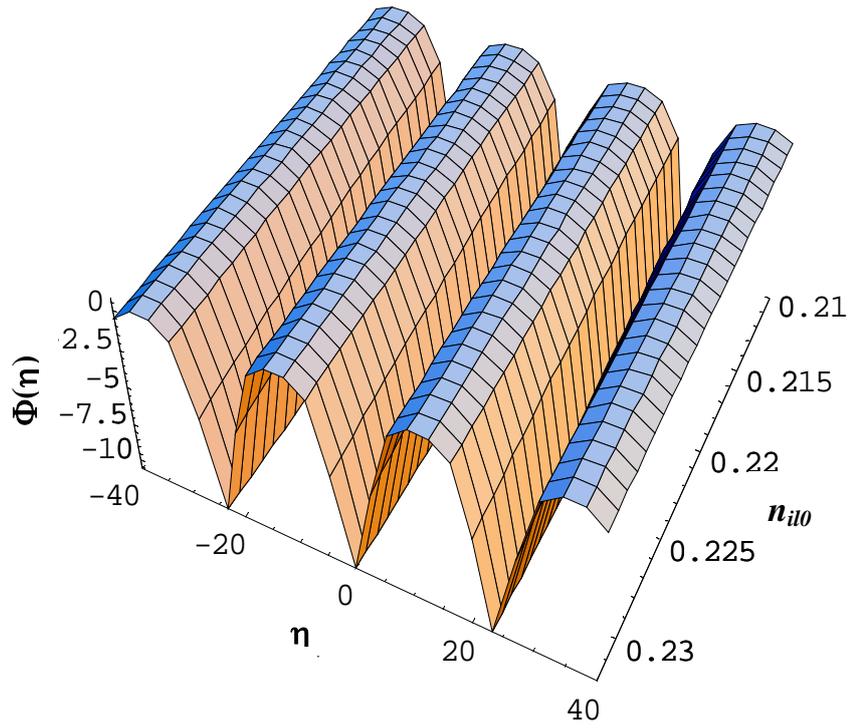

(a)

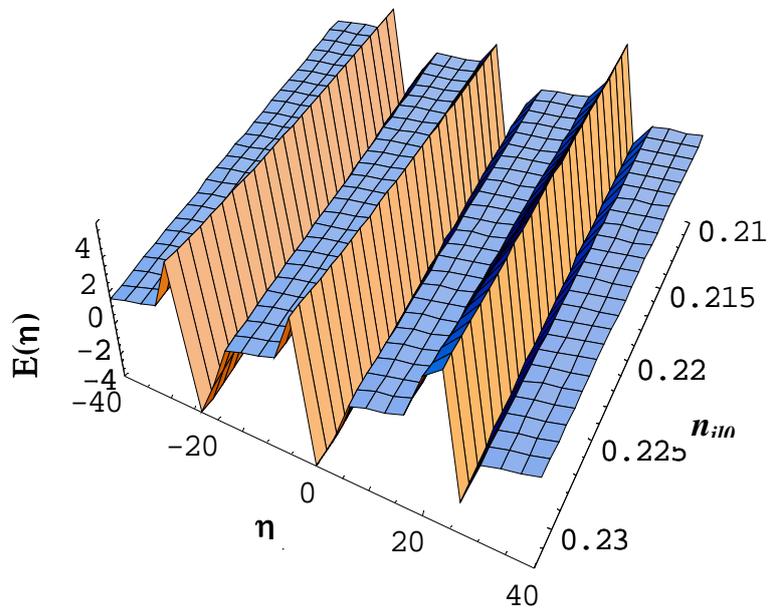

(b)

Fig. (8)